\documentclass[conference]{IEEEtran}
\IEEEoverridecommandlockouts
\usepackage{amsmath,amssymb,amsfonts}
\usepackage{graphicx}
\usepackage{booktabs}
\usepackage{bm}
\usepackage{array}
\usepackage{tikz}
\usepackage{tabularx}
\usepackage{pgfplots}
\usepackage{stfloats}
\usepackage{xcolor}
\pgfplotsset{compat=1.18}
\usetikzlibrary{positioning,arrows.meta,fit,backgrounds,calc,shapes.geometric,shadows}

\makeatletter
\g@addto@macro\normalsize{%
  \setlength\abovedisplayskip{4pt plus 2pt minus 1pt}%
  \setlength\belowdisplayskip{4pt plus 2pt minus 1pt}%
  \setlength\abovedisplayshortskip{0pt plus 2pt}%
  \setlength\belowdisplayshortskip{2pt plus 2pt minus 1pt}}
\makeatother

\begin{document}
\title{Resilient Cell-Free MIMO ISAC Under Uncertainties: An Impairment-Aware Graph Neural Network Framework}
\author{\IEEEauthorblockN{Oways Alsoloh, Robert-Jeron Reifert, Aymen Khaleel, Stefan Roth and Aydin Sezgin}
\IEEEauthorblockA{Institute of Digital Communication Systems, Ruhr University Bochum,
Bochum, Germany\\
Email: \{alsoloh.oways, robert-.reifert, aymen.khaleel, stefan.roth, aydin.sezgin\}@rub.de}
\thanks{This work was supported in part by the German Research Foundation (“Deutsche Forschungsgemeinschaft”) (DFG) under Project–ID 287022738 TRR 196 for Project S02 and in part by the German Federal Ministry of Research, Technology and Space (BMFTR) in the course of the 6GEM+
Transfer Hub under grant 16KIS2411.}}

\maketitle
\begin{abstract} Beamforming designs for integrated sensing and communication (ISAC) in cell-free multiple-input multiple-output (MIMO) systems face the challenge to stay reliable under uncertainties from both channel state information (CSI) as well as clutter. To this end, this paper proposes an impairment-aware heterogeneous graph neural network (GNN) framework for beamforming that obtains the CSI error level and the clutter covariance as edge features, which enables a single trained model to adapt to both impairments. In particular, the GNN is trained on a low quantile of the minimum user rate with a clutter-aware signal-to-clutter-plus-noise ratio, balanced by a normalized Tchebycheff scalarization. Simulation results show that the proposed GNN framework outperforms the considered baselines in sensing-only and communication-only settings. Depending on the operating point, a single GNN model attains approximately~97\% of each baseline's rate while serving both functionalities at once. The model further generalizes to unseen CSI error levels and clutter counts and renders the ISAC system resilient. \looseness=-1
\end{abstract}
\begin{IEEEkeywords}
Cell-free, MIMO, integrated sensing and communication, GNN, imperfect CSI, clutter awareness, resilience.
\end{IEEEkeywords}
%==============================================================================
\section{Introduction}
%==============================================================================
Integrated sensing and communication (ISAC) describes the simultaneous utilization of the same radio hardware for communication and sensing, and is considered an important technology for future 6G systems~\cite{liu2022,luo2026}. Cell-free multiple-input multiple-output (MIMO) is an attractive architecture for ISAC, since its distributed access points (APs) provide macro diversity to the user equipments (UEs) and observe the target from several positions~\cite{demirhan2023}. Still, this diversity alone does not guarantee reliable communication, since the beamformers are computed based on imperfect channel state information (CSI). In particular, beamformer designs that treat the channel estimates as exact leave residual interference that grows with the estimation error. Further, a beamformer designed for good average rates can lead to outages, in which weak users fall far below their target rate in a non-negligible fraction of channel realizations. This outcome motivates optimizing a low quantile of the minimum user rate instead~\cite{liang2024}. For sensing, similar problems arise. Here, clutter reflectors create undesired echoes that compete with the sensing target (ST) echo. Since clutter arrives from directions that the receiving AP cannot fully separate from the target direction, part of the clutter power cannot be suppressed, which reduces the signal-to-clutter-plus-noise ratio (SCNR). This makes sensing performance dependent not only on target illumination, but also on how the beamformers interact with the spatial clutter structure~\cite{dassanayake2025}. Thus, a robust ISAC design must account for both CSI uncertainty in communication and clutter in sensing. Moreover, sensing and communication affect each other, since they share the same transmit resources~\cite{umra2026}. As no beamformer maximizes both functionalities at once, balancing these priorities needs to be addressed. A common way to select a suitable operating point is to maximize a weighted sum of the robust communication and sensing rates. However, the two rates differ in scale, which makes the effect of a given weight hard to control. The normalized Tchebycheff formulation adopted in this work instead measures each rate relative to its own reference value~\cite{meng2024}. Yet, the resulting problem is non-convex and would have to be solved anew for every channel and clutter realization, which motivates a learning-based mapping from the available system information to the beamformers~\cite{reifert2025}. \looseness=-1

Graph neural networks (GNNs) are well suited for this mapping, since a cell-free ISAC system can be represented directly as a graph. The APs, UEs and ST form the nodes, while the AP-UE and AP-ST links form the edges. This representation offers two advantages. First, a GNN shares its weights across all nodes of the same type, which makes the beamformer independent of the ordering of APs and UEs and keeps the number of trainable parameters independent of the number of APs and UEs. Second, the impairments considered here belong to the individual links, namely the CSI error to the AP-UE links and the clutter to the AP-ST links. Therefore, they can be passed to the GNN directly as edge features~\cite{demirhan2024}.\looseness=-1

Existing cell-free ISAC designs address CSI uncertainty, clutter, and the communication-sensing trade-off only partially. A cell-free ISAC beamforming problem is solved in~\cite{demirhan2023} by iterative convex methods under perfect CSI and a clutter-free target. In~\cite{demirhan2024,sacgnn2024}, cell-free ISAC beamforming is learned by heterogeneous GNNs that exploit the AP-UE-target topology to maximize an average-rate objective with a sensing signal-to-noise ratio (SNR) term. Outage-oriented robustness to CSI error is addressed in~\cite{liang2024} through rate-quantile maximization with uncertainty sampling, but limited to a single-transmitter multiple-input single-output downlink without sensing. Clutter and imperfect channel estimation are modeled jointly in~\cite{dassanayake2025}, where power allocation is optimized over fixed beam directions under an ergodic-rate criterion. None of the prior designs learns full ISAC beamformers under clutter and imperfect CSI with an outage-oriented criterion, nor examines resilience to uncertainties, which is the regime targeted in this paper. \looseness=-1

The main contribution of this work is a heterogeneous GNN that receives a per-AP clutter covariance and the CSI-error level as edge features and maps them, together with the channel estimates and target steering vectors, directly to the beamformers. The GNN is trained on an outage-oriented quantile of the minimum user rate and a clutter-aware SCNR, combined by a normalized Tchebycheff scalarization whose weight traces the communication-sensing trade-off. Beyond robustness to the impairments seen during training, the impairment-aware inputs make a single trained model resilient at CSI-error levels and clutter counts outside the training range. Here, resilience refers to maintaining the system functionality under unforeseen impairments, of which robustness is one aspect~\cite{reifert2023}. \looseness=-1

\textit{Notation:} Bold uppercase and lowercase letters denote matrices and
column vectors, respectively. $(\cdot)^{T}$ and $(\cdot)^{H}$ denote the
transpose and Hermitian transpose, $\|\cdot\|$ the Euclidean norm, $|\cdot|$
the absolute value, and $\Re(\cdot)$ and $\Im(\cdot)$ the real and imaginary
parts. $\mathbf I_N$ is the $N\times N$ identity matrix, $\delta_{uu'}$ the
Kronecker delta, and $\mathrm{vec}(\cdot)$ stacks the columns of a matrix into
a vector. $\mathbb E[\cdot]$ and $\Pr(\cdot)$ denote expectation and
probability, $\mathcal{CN}(\mathbf 0,\boldsymbol\Sigma)$ the zero-mean circularly
symmetric complex Gaussian distribution with covariance $\boldsymbol\Sigma$, and $\mathrm{Unif}[a,b)$ the uniform
distribution on $[a,b)$. Finally, $[x]_{+}=\max(x,0)$ denotes the positive
part.\looseness=-1
%==============================================================================
\section{System Model and Problem Formulation}
%==============================================================================
We consider a downlink cell-free ISAC system with \(M\) distributed APs and \(K\) single-antenna communication users, each served by a single data stream. The AP set is denoted as \(\mathcal M\). Each AP is equipped with a transmit and a receive uniform linear array (ULA), each consisting of $N$ antennas. All APs are coordinated by a central processing unit and simultaneously serve the users while illuminating the target embedded in a cluttered scene, using the same time and frequency resources. The full setup is illustrated in Fig.~\ref{fig:systemmodel}.\looseness=-1

\subsection{Signal Model}
% ---------- figure ----------
% ---- icon macros (place once; currently sit just before the figure) ----
\newcommand{\apicon}[2]{%
  \begin{scope}[shift={(#1,#2)},line width=0.9pt]
    \draw (-0.32,-0.45) -- (-0.10,0.32) -- (0.10,0.32) -- (0.32,-0.45);
    \draw (-0.24,-0.18) -- (0.24,-0.18);
    \draw (-0.17,0.05) -- (0.17,0.05);
    \draw (-0.10,0.32) -- (-0.20,0.55);
    \draw ( 0.10,0.32) -- ( 0.20,0.55);
    \draw (0,0.32) -- (0,0.58);
    \draw (-0.26,0.70) arc (150:30:0.30);
  \end{scope}}
\newcommand{\ueicon}[2]{%
  \draw[line width=0.9pt,rounded corners=1.5pt]
    (#1-0.17,#2-0.30) rectangle (#1+0.17,#2+0.30);
  \draw[line width=0.9pt] (#1-0.10,#2+0.21) -- (#1+0.10,#2+0.21);}
\newcommand{\caricon}[2]{%
  \begin{scope}[shift={(#1,#2)},line width=0.8pt]
    \fill[rounded corners=2pt] (-0.40,-0.10) rectangle (0.40,0.11);
    \fill[rounded corners=2pt] (-0.21,0.09) rectangle (0.19,0.27);
    \fill (-0.25,-0.16) circle (0.09);
    \fill ( 0.25,-0.16) circle (0.09);
  \end{scope}}
\newcommand{\clicon}[2]{%
  \begin{scope}[shift={(#1,#2)}]
    \fill[orange!75!black]
      (-0.30,-0.18) -- (-0.12,0.22) -- (0.10,-0.02) -- (0.30,0.20) -- (0.26,-0.18) -- cycle;
  \end{scope}}
\newcommand{\cpuicon}[2]{%
  \begin{scope}[shift={(#1,#2)},line width=0.8pt]
    \foreach \k in {0,1,2}{\fill (-0.32,-0.48+\k*0.36) rectangle (0.32,-0.26+\k*0.36);}
  \end{scope}}

\tikzset{
  comm/.style={-{Stealth[length=2.4mm]}, line width=1.0pt, blue!70!black},
  sensbi/.style={{Stealth[length=2.4mm]}-{Stealth[length=2.4mm]},
                 line width=1.0pt, red!80!black},
  clutbi/.style={{Stealth[length=2.4mm]}-{Stealth[length=2.4mm]},
                 line width=1.0pt, orange!85!black, dashed},
  bh/.style={green!50!black, line width=1.0pt, dash pattern=on 3pt off 2.5pt},
  lbl/.style={font=\small}, leg/.style={font=\small}
}
\newcommand{\legline}[1]{\tikz[baseline=-0.55ex]{\draw[#1] (0,0) -- (0.55,0);}}
\newcommand{\legsep}{\hspace{1.6em}}

\begin{figure}[t]
\centering
\resizebox{\columnwidth}{!}{%
\begin{tikzpicture}
% extends (never clips) the bounding box, and is symmetric about the drawing centre
\path (-1.75,-3.78) rectangle (9.29,3.05);

%---- regions ----
\begin{scope}[on background layer]
  \fill[blue!11]  (0.7, 1.35) rectangle (9.1, 2.70);
  \fill[blue!11]  (0.7,-2.70) rectangle (9.1,-1.35);
  \fill[green!13] (1.3,-0.72) rectangle (8.6, 0.98);
\end{scope}
\node[lbl,blue!55!black]  at (5.30,2.35) {$\mathcal{M}$};
\node[lbl,green!40!black] at (1.60,0.72) {$\mathcal{K}$};

%---- fronthaul ----
\draw[bh] (8.5, 2.92) -- (0.25, 2.92) -- (0.25,-2.92) -- (8.5,-2.92);
\draw[bh] (-0.85, 0.10) -- (0.25, 0.10);
\draw[bh] (0.25,  1.80) -- (1.18,  1.80);
\draw[bh] (0.25, -1.80) -- (1.18, -1.80);
\draw[bh] (8.50,  2.92) -- (8.50,  2.58);
\draw[bh] (8.50, -2.92) -- (8.50, -2.32);

%---- nodes ----
\apicon{1.6}{1.80}   \node[lbl] at (2.15, 2.00)  {AP};
\apicon{8.5}{1.80}   \node[lbl] at (8.95, 1.15)  {AP};
\apicon{1.6}{-1.80}  \node[lbl] at (1.60,-2.45)  {AP};
\apicon{8.5}{-1.80}  \node[lbl] at (8.05,-2.50)  {AP};
\ueicon{2.6}{0.10}   \node[lbl] at (3.10, 0.10)  {UE};
\ueicon{6.4}{0.42}   \node[lbl] at (5.75, 0.42)  {UE};
\ueicon{7.9}{-0.28}  \node[lbl] at (7.90, 0.32)  {UE};
\caricon{4.3}{0.28}  \node[lbl] at (4.30,-0.44)  {ST};
\clicon{3.30}{1.12}  \node[lbl,orange!70!black] at (3.30, 1.62) {Clutter};
\clicon{7.05}{-1.05} \node[lbl,orange!70!black] at (7.05,-0.56) {Clutter};
\cpuicon{-1.4}{0.10} \node[lbl] at (-1.40, 0.85) {CPU};

%---- communication ----
\draw[comm] (1.85, 1.38) -- (2.50, 0.45);
\draw[comm] (8.28, 1.38) -- (6.62, 0.64);
\draw[comm] (1.85,-1.38) -- (2.50,-0.24);
\draw[comm] (8.32,-1.38) -- (7.98,-0.62);

%---- target illumination and echo ----
\draw[sensbi] (4.02, 0.52) -- (1.95, 1.42);
\draw[sensbi] (4.65, 0.52) -- (8.18, 1.48);
\draw[sensbi] (4.02, 0.10) -- (1.95,-1.42);
\draw[sensbi] (4.58, 0.08) -- (8.20,-2.05);

%---- clutter illumination and echo ----
\draw[clutbi] (3.02, 1.22) -- (2.02, 1.55);
\draw[clutbi] (7.35,-1.18) -- (8.15,-1.55);

%---- legend (centred single row) ----
\node[leg] at (3.77,-3.48) {%
  \legline{sensbi}\,target echo\legsep
  \legline{comm}\,comm. link\legsep
  \legline{clutbi}\,clutter echo\legsep
  \legline{bh}\,fronthaul};

\end{tikzpicture}}
\caption{Cell-free MIMO ISAC scenario, with distributed APs serving the UEs and sensing a target in the presence of clutter.}
\label{fig:systemmodel}
\vspace{-4mm}
\end{figure}
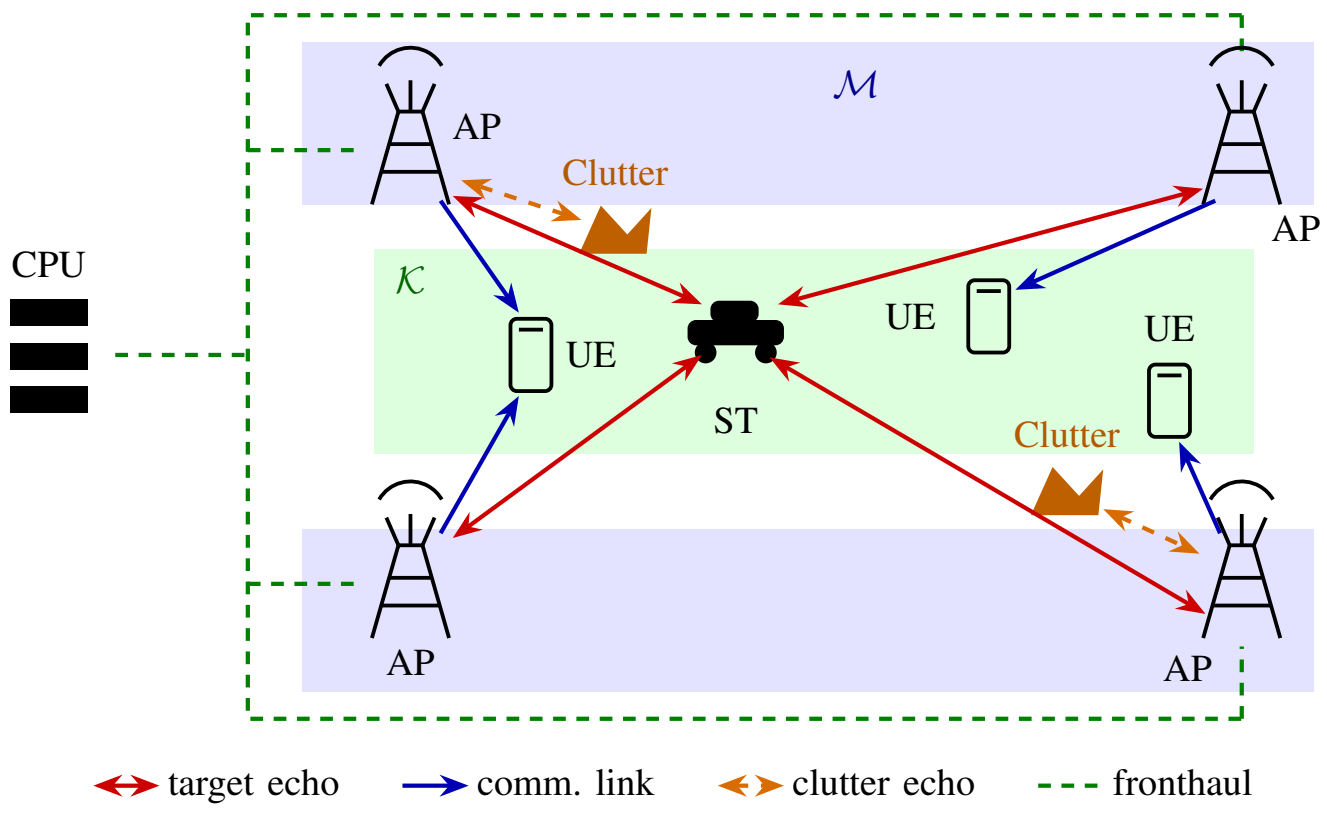
The signal $\mathbf{x}_m\in\mathbb{C}^{N\times 1}$ transmitted from the antenna array of the $m$-th AP combines the data symbol $x_k$ intended for user $k\in\mathcal{K}$ and dedicated sensing signal $x_s$ for $s\in\mathcal{S}$, where $\mathcal{K}$ and $\mathcal{S}$ are the sets of communication and sensing streams, respectively, with $\mathcal{K}\cap\mathcal{S}=\emptyset$. For compactness, the overall set of both streams is denoted as $\mathcal{U}=\mathcal{K} \cup \mathcal{S}$. The transmitted signal from the $m$-th AP is expressed as \cite{demirhan2023}
\begin{align}
\mathbf{x}_m =\sum_{k \in \mathcal{K}}\mathbf{w}_{mk}\,x_k
+\sum_{s \in \mathcal{S}}\mathbf{w}_{ms}\,x_s =\sum_{u \in \mathcal{U}}\mathbf{w}_{mu}\,x_u,
\label{eq:txsignal}
\end{align}
with $\mathbf{w}_{mu}\in \mathbb{C}^{N\times 1}$ the beamforming vector of the $u$-th stream at AP $m$. The streams are mutually uncorrelated with unit power, $\mathbb{E}[x_u x_{u'}^{*}]=\delta_{uu'}$, and each AP satisfies a total transmit power
constraint given by the sum of the squared norms of its beamformers,
\begin{equation}
\mathbb{E}\big[\|\mathbf{x}_m\|^2\big]
=\sum_{u\in\mathcal{U}}\|\mathbf{w}_{mu}\|^2\leq P_m.
\label{eq:power}
\end{equation}
Since all APs transmit the same symbol $x_u$, the per-AP beamformers of
stream $u$ can be stacked into a single vector $\mathbf{w}_u=~\left[\mathbf{w}_{1u}^{T},\dots,
\mathbf{w}_{M u}^{T}\right]^{T}\in\mathbb{C}^{M N\times 1}$, which is denoted by $\mathbf w_k$ for a communication stream $k$ and by $\mathbf w_s$ for a sensing stream $s$.\looseness=-1

\subsection{Communication Model} The communication channel between AP $m$ and user $k$ is denoted as $\mathbf{h}_{mk}\in\mathbb{C}^{N\times 1}$. By stacking the channel vectors associated with all APs, we obtain the aggregated channel vector $\mathbf{h}_k\in\mathbb{C}^{M N\times 1}$, which characterizes the communication channel between user $k$ and the entire set of APs. Hence, the true channel is written as the sum of the channel estimate $\tilde{\mathbf{h}}_k$ available at the APs and an independent error vector $\mathbf{e}_k$ as \cite{liang2024}
\begin{equation}
\mathbf{h}_k=\tilde{\mathbf{h}}_k+\mathbf{e}_k,\qquad
\mathbf{e}_k\sim\mathcal{CN}(\bm 0,\sigma_e^2\mathbf{I}_{MN}),
\label{eq:csierror}
\end{equation}
where $\tilde{\mathbf{h}}_k \sim\mathcal{CN}(\bm 0, \mathbf{I}_{MN})$ is modeled as uncorrelated Rayleigh fading channel. Under the block-fading assumption, the received signal at the $k$-th UE is
\begin{equation}
y_k=\sum_{m\in\mathcal{M}}\mathbf{h}_{mk}^{H}\mathbf{x}_m
+n_k,
\label{eq:rxk}
\end{equation}
with $n_k\sim\mathcal{CN}(0,\sigma_k^2)$ being the additive white Gaussian noise with power $\sigma_k^2$.
The received signal separates into the desired stream, the multi-user interference, the sensing interference, and the noise. Based on this decomposition, the signal-to-interference-plus-noise ratio (SINR) of the communication signal received at the $k$-th UE follows as \looseness=-1 \cite{demirhan2023}
\begin{equation}
\Gamma_k^{(\mathrm{c})}=\frac{|\mathbf{h}_k^{H}\mathbf{w}_k|^2}
{\sum\nolimits_{k'\in\mathcal{K}\setminus\{k\}}
|\mathbf{h}_k^{H}\mathbf{w}_{k'}|^2
+\sum\nolimits_{s\in\mathcal{S}}|\mathbf{h}_k^{H}\mathbf{w}_s|^2
+\sigma_k^2}.
\label{eq:sinr}
\end{equation}
Subsequently, the corresponding user rate is
\begin{equation}
R_k^{(\mathrm{c})}=\log_2(1+\Gamma_k^{(\mathrm{c})}).
\label{eq:rate}
\end{equation}
\subsection{Sensing Model}
A distributed monostatic sensing model is adopted, in which each AP processes only its own echo. This avoids the tight inter-AP synchronization required by multistatic processing, limits the fronthaul load, and makes the clutter seen by each AP a local quantity that can be attached to its sensing edge in Section~\ref{sec:method}. Each AP operates in full-duplex mode with co-located transmit and receive arrays, and the transmit-to-receive self-interference is assumed to be cancelled prior to sensing processing, which is feasible through combined analog and digital cancellation~\cite{barneto2019}. The signal transmitted by AP $m$ returns after a single bounce from a point-like target and $C$ undesired clutter reflectors \cite{dassanayake2025}. The target gain $\alpha_m\sim\mathcal{CN}(0,\zeta_m^2)$ captures the path loss and the radar cross section (RCS), whereas $\alpha_{m,c}\sim\mathcal{CN}(0,\sigma_{\alpha}^2)$ denotes the reflection coefficient of the $c$-th clutter source. Let $\mathbf a(\phi)$ denote the array response vector with unit-modulus entries $\|\mathbf a(\phi)\|^2=N$. $\phi_{m}$ denotes the angle of the target, and $\phi_{m,c}$ the angle of the $c$-th clutter reflector seen by AP $m$. The sensing channel is
given by \looseness=-1
\begin{equation}
\mathbf{G}_{m}=\alpha_{m}\mathbf{a}(\phi_{m})\mathbf{a}^{H}(\phi_{m})
+\sum_{c=1}^{C}\alpha_{m,c}\mathbf{a}(\phi_{m,c})\mathbf{a}^{H}(\phi_{m,c}).
\label{eq:Gsense}
\end{equation}
Under the Swerling-I model, the reflection gains are constant over the coherent processing interval and vary independently across intervals. With this model, the echo received at AP $m$ is
\begin{equation}
\mathbf{y}_{m}=\mathbf{G}_{m}\,\mathbf{x}_{m}+\mathbf{n}_{m},
\label{eq:radarrx}
\end{equation}
where $\mathbf{n}_{m}\sim\mathcal{CN}(\bm 0,\varsigma^2_{m}\mathbf{I}_{N})$ is the radar receiver noise with the power $\varsigma^2_{m}$. Each AP $m$ applies the normalized target-matched combiner $\mathbf{a}^H(\phi_{m})/N$ to its own echo. Since the target gains are independent across APs, the squared magnitudes of the combiner outputs are aggregated across the APs~\cite{demirhan2023}. Clutter directions are generally not orthogonal to the target direction, and part of the clutter therefore leaks through the matched combiner. The normalized target-clutter correlation at AP \(m\) is expressed as~$\kappa_{m,c} = \frac{\mathbf{a}^H(\phi_{m})\mathbf a(\phi_{m,c})}{N},$ where the absolute value of $\kappa_{m,c}$ is bounded by zero and one. For a given clutter realization, we define the effective clutter covariance matrix at AP \(m\) as
\begin{equation}
\mathbf Q_{\mathrm{cl},m} = \sum_{c=1}^{C} |\alpha_{m,c}|^2 |\kappa_{m,c}|^2 \mathbf a(\phi_{m,c})\mathbf{a}^H(\phi_{m,c}). \label{eq:clutter_cov}
\end{equation}
Subsequently, the clutter power leakage through the target-matched combiner is given by
\begin{equation}
I_{\mathrm{cl}} = \sum_{m\in\mathcal M} \sum_{u\in\mathcal U} \mathbf w_{mu}^{H}\mathbf Q_{\mathrm{cl},m}\mathbf w_{mu}. \label{eq:clutter_power}
\end{equation}
The echo power is averaged over the Swerling-I target gain, while the clutter coefficients and angles $\{\phi_{m,c},\alpha_{m,c}\}$ are fixed. Consequently, we define the aggregate SCNR as the ratio of the total target echo power to the total clutter and noise power to assess sensing performance using observations from all APs
\begin{equation}
\Gamma^{(\mathrm s)}
=\frac{\sum_{m\in\mathcal M}\zeta_{m}^{2}\sum_{u\in\mathcal U}
\left|\mathbf{a}^H(\phi_{m})\mathbf w_{mu}\right|^2}
{I_{\mathrm{cl}}+\bar\varsigma^2},
\label{eq:scnr}
\end{equation} 
with the normalized noise power \(\bar\varsigma^2 =\sum_{m\in\mathcal M}\varsigma_{m}^2/N\). The sensing performance is then expressed as
\begin{equation}
R^{(\mathrm s)}=\log_2\!\left(1+\Gamma^{(\mathrm s)}\right).
\label{eq:sensrate}
\end{equation}
For a Gaussian target gain observed at a single AP, maximizing this rate is equivalent to minimizing the mean squared error (MSE) of the gain estimate~\cite{yang2007}.
%==============================================================================
\subsection{Problem Formulation} \label{sec:prob}
%==============================================================================
The minimum user rate is denoted by $R_{\min}^{(\mathrm c)}=\min_{k\in\mathcal K}R_k^{(\mathrm c)}$, and $r^{(\mathrm c)}$ denotes the reliable communication rate level associated with the outage probability $\rho$. The sensing objective is the sensing rate $R^{(\mathrm s)}$. The minimum over the users ensures fairness, which a single target does not require. The two objectives generally conflict, since the communication and sensing streams share the same transmit power budget and the sensing streams contribute interference to the communication users as mentioned in~\eqref{eq:sinr}. Therefore, a multi-objective optimization problem (MOOP) is formulated as
\begin{subequations}\label{eq:moop}
\begin{align}
\max_{\{\mathbf w_{mu}\}}\quad
&
\left(r^{(\mathrm c)},\ R^{(\mathrm s)}\right) \tag{\ref*{eq:moop}}\\
\text{s.t.}\quad
& \Pr\!\left(R_{\min}^{(\mathrm c)}\le r^{(\mathrm c)}\right)\le\rho, \label{eq:moopa} \\
& \sum_{u\in\mathcal U}\|\mathbf w_{mu}\|^2\le P_m, && \forall m\in\mathcal M. \label{eq:moopb}
\end{align}
\end{subequations}
Constraint~\eqref{eq:moopa} means that the minimum user rate falls below $r^{(\mathrm c)}$ with a probability no larger than $\rho$. Constraint~\eqref{eq:moopb} is the per-AP transmit power budget. The MOOP is non-convex due to the coupled SINR and SCNR expressions, while the outage constraint has no closed-form expression under imperfect CSI. The non-convexity and the missing closed form of the outage constraint prevent a direct solution with standard convex solvers. \looseness=-1
%==============================================================================
\section{Proposed Robust Learning Method} \label{sec:method}
%==============================================================================
To address the MOOP, we propose a learning framework illustrated in Fig.~\ref{fig:method}. First, we introduce a CSI error- and clutter-aware GNN that maps the available information to the beamformers. Subsequently, the training objective is presented. 
\subsection{CSI Error- and Clutter-Aware GNN}
\begin{figure}[t]
\centering
\begin{tikzpicture}[
  font=\scriptsize, >={Stealth[length=1.4mm]},
  box/.style={draw, rounded corners=1.5pt, align=center, inner sep=1.8pt,
              fill=white, minimum height=4.6mm},
  ap/.style={draw, circle, fill=gray!25, minimum size=2.4mm, inner sep=0pt},
  ue/.style={draw, circle, fill=blue!15, minimum size=2.2mm, inner sep=0pt},
  tg/.style={draw, circle, fill=red!25, minimum size=2.2mm, inner sep=0pt},
  cedge/.style={blue!60!black, line width=0.3pt},
  sedge/.style={red!70!black, line width=0.7pt},
  ccol/.style={blue!60!black}, scol/.style={red!70!black},
]
%---------------- graph with edge features --------------------------------
\coordinate (T) at (0.75,0.72);
\foreach \i/\y in {1/1.17,2/0.72,3/0.27} \coordinate (A\i) at (2.35,\y);
\foreach \k/\y in {1/1.25,2/0.87,3/0.49,4/0.11} \coordinate (U\k) at (4.15,\y);
\foreach \i in {1,2,3}{
  \draw[sedge] (A\i) -- (T);
  \foreach \k in {1,2,3,4} \draw[cedge] (A\i) -- (U\k);}
\node[tg] at (T) {};
\foreach \i in {1,2,3} \node[ap] at (A\i) {};
\foreach \k in {1,2,3,4} \node[ue] at (U\k) {};
\node[below=0.9mm] at (T) {ST};
\node[below=0.9mm] at (A3) {APs};
\node[below=0.9mm] at (U4) {UEs};
\node[scol] at (1.4,1.35) {$\mathbf a(\phi_m),\,\mathbf Q_{\mathrm{cl},m}$};
\node[ccol] at (3.3,1.45) {$\tilde{\mathbf h}_{mk},\,\sigma_e$};
\begin{scope}[on background layer]
  \node[draw, rounded corners=2pt, fill=white,
        double copy shadow={shadow xshift=0.9mm, shadow yshift=0.9mm},
        fit={(0.2,-0.4) (4.85,2.0)}, inner sep=0pt] (G) {};
\end{scope}
\path (0,2.25);
\node[anchor=north, inner sep=1.5pt] at (G.north)
  {Message-passing layer $\ell=1,\dots,L$};
%---------------- readout -------------------------------------------------
\node[box, text width=1.7cm, fill=gray!8] (w) at (6.62,0.8) {$\{\mathbf w_{mu}\}$};
\draw[->] ([xshift=1.8mm]G.east |- w) -- (w.west);
%---------------- training ------------------------------------------------
\node[box, fill=blue!6, text width=3.3cm] (ce) at (6.3,-1.02)
  {$\hat r^{(\mathrm c)}$ \eqref{eq:emp_comm_quantile}};
\node[box, fill=red!6, text width=3.3cm] (se) at (6.3,-1.72)
  {$\hat r^{(\mathrm s)}$ \eqref{eq:emp_sensing_rate}};
\node[box, text width=2.2cm] (L) at (2.3,-1.3)
  {Tchebycheff loss $\mathcal L$};
% bus from beamformers to both evaluations
\coordinate (bus) at ([xshift=2.2mm]ce.east |- w.east);
\draw (w.east) -- (bus);
\draw[->] (bus) |- (ce.east);
\draw[->] (bus) |- (se.east);
\fill (bus |- ce.east) circle (0.5pt);
\draw[->, ccol] (ce.west) -- ([yshift=1.1mm]L.east);
\draw[->, scol] (se.west) -- ([yshift=-1.1mm]L.east);
\draw[->, dashed, thick, gray!70!black] (L.north) --
  node[right, inner sep=1pt]{$\nabla\mathcal L$} (L.north |- G.south);
\begin{scope}[on background layer]
  \fill[gray!12, rounded corners=2pt] (0.0,-2.05) rectangle (8.4,-0.52);
\end{scope}
\node[anchor=south west, inner sep=1.5pt, gray!40!black] at (0.0,-2.05) {Training};
\end{tikzpicture}
\caption{Proposed impairment-aware GNN framework, which maps the clutter-aware and CSI error-aware edge features to the beamformers $\{\mathbf w_{mu}\}$, trained with the Tchebycheff loss of~\eqref{eq:emp_comm_quantile} and~\eqref{eq:emp_sensing_rate}.}
\label{fig:method}
\vspace{-4mm}
\end{figure}
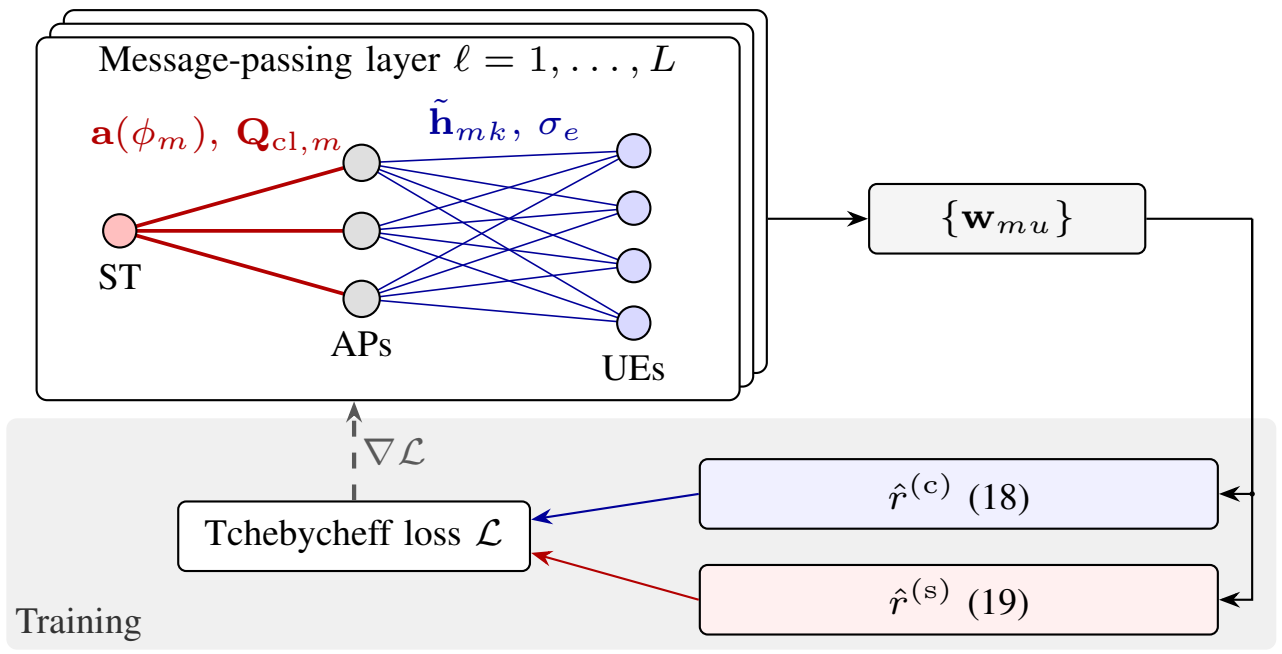

The beamformer is parameterized by a heterogeneous GNN whose graph follows the AP-UE-ST structure of \cite{demirhan2024}, with the AP-UE edges carrying communication channel features and the AP-ST edges carrying sensing and clutter features. The GNN represents each node and edge by an embedding, which is a numerical vector that summarizes its associated information. Let $\mathbf{ap}_m^{(\ell)}$, $\mathbf{ue}_k^{(\ell)}$, and $\mathbf{st}^{(\ell)}$ denote the node embeddings at layer $\ell$, and $\mathbf{ce}_{mk}^{(\ell)}$, $\mathbf{se}_{m}^{(\ell)}$ the communication and sensing edge embeddings. The input maps $\mathrm{Emb}_{\mathrm c}$ and $\mathrm{Emb}_{\mathrm s}$ are linear embedding layers, whereas the edge-update, node-update and readout maps are multi-layer perceptrons (MLPs) with LeakyReLU activations. Each communication edge is initialized from the channel estimate together with the error level that characterizes its accuracy, 
\begin{equation}
\mathbf{ce}_{mk}^{(0)}
=
\mathrm{Emb}_{\mathrm c}
\big(
\big[
\Re\big(\tilde{\mathbf h}_{mk}\big)^{T},
\Im\big(\tilde{\mathbf h}_{mk}\big)^{T},
\sigma_e
\big]^{T}
\big),
\label{eq:comm_edge_init}
\end{equation}
which allows a single trained model to be operated across CSI-error levels. With the target and clutter blocks 
\begin{align}
  \boldsymbol{\eta}_m^{\mathrm{tg}} &=
    \big[
    \Re\big(\mathbf a(\phi_{m})\big)^{T},
    \Im\big(\mathbf a(\phi_{m})\big)^{T}
    \big]^{T},\nonumber \\  
    \boldsymbol{\eta}_m^{\mathrm{cl}} &=
    \big[
    \Re\big(\mathrm{vec}(\mathbf Q_{\mathrm{cl},m})\big)^{T},
    \Im\big(\mathrm{vec}(\mathbf Q_{\mathrm{cl},m})\big)^{T}
    \big]^{T}, \nonumber 
\end{align}
the sensing edge associated with AP $m$ is initialized as
\begin{equation}
\mathbf{se}_{m}^{(0)} = \mathrm{Emb}_{\mathrm s} \big( \big[ {\boldsymbol{\eta}_m^{\mathrm{tg}}}^{T}, {\boldsymbol{\eta}_m^{\mathrm{cl}}}^{T}\big]^{T} \big). \label{eq:sens_edge_init}
\end{equation} \looseness=-1
Thus, the sensing branch receives the clutter covariance that determines the leakage in~\eqref{eq:clutter_power}. At message-passing layer $\ell$, the edge embeddings are updated as \looseness=-1
\begin{align}
\mathbf{ce}_{mk}^{(\ell)}
&=
f_{\mathrm c}^{(\ell-1)}
\left(
\mathbf{ce}_{mk}^{(\ell-1)},\mathbf{ce}_{mk}^{(0)},
\mathbf{ap}_{m}^{(\ell-1)},\mathbf{ue}_{k}^{(\ell-1)}
\right),\\
\mathbf{se}_{m}^{(\ell)}
&=
f_{\mathrm s}^{(\ell-1)}
\left(
\mathbf{se}_{m}^{(\ell-1)},\mathbf{se}_{m}^{(0)},
\mathbf{ap}_{m}^{(\ell-1)},\mathbf{st}^{(\ell-1)}
\right),
\end{align}
where the initial edge feature is reinjected at every layer, keeping the impairment state available throughout the network. Fig.~\ref{fig:method} depicts one layer, which is repeated $L$ times. The updated edge embeddings are averaged at each node, and each node type updates its embedding with its own MLP. After $L$ layers, readout MLPs map the final edge embeddings to unconstrained beamformers, which a projection step scales to satisfy~\eqref{eq:power}. This yields the beamformers $\{\mathbf w_{mu}\}$ in Fig.~\ref{fig:method}, comprising $\mathbf w_{mk}$ and $\mathbf w_{ms}$ for the communication and sensing beamformers, respectively.

%-----------------------------------------------------------------
\subsection{Normalized Tchebycheff Loss}
The MOOP is scalarized by a Tchebycheff function~\cite{miettinen1999}, which is defined with respect to a reference point that expresses the desired operating level of each objective. The GNN is trained in an unsupervised manner. Both communication and sensing metrics are estimated empirically over each mini-batch of size $B$. To account for imperfect CSI, the reliable communication rate is estimated through data augmentation, following \cite{liang2024,reifert2024}. For each batch sample $b$, we generate $J$ channel realizations for every user $k$ by adding independent error vectors to the channel estimate $\tilde{\mathbf h}_{k,b}$,
\[
\mathbf h_{k,b}^{(j)}
=
\tilde{\mathbf h}_{k,b}+\mathbf e_{k,b}^{(j)},
\qquad j=1,\ldots,J.
\]
The minimum user rate $R_{\min,b,j}^{(\mathrm c)}$ is computed for each realization $j$, and the lower-tail value $r_b^{(\mathrm c)}$ is taken as the empirical $\rho$-quantile of the resulting samples. Averaging over the mini-batch gives
\vspace{-1mm}
\begin{equation}
\hat r^{(\mathrm c)}
=
\frac{1}{B}\sum_{b=1}^{B} r_b^{(\mathrm c)} .
\label{eq:emp_comm_quantile}
\end{equation}
The beamformers are designed from the available estimates alone, whereas the
rates entering~\eqref{eq:emp_comm_quantile} are evaluated on the sampled channel realizations $\mathbf h_{k,b}^{(j)}$, which emulate the unknown true channels. The GNN therefore learns beamformers that remain reliable under the estimation error. The sensing value is the mini-batch average\looseness=-1
\vspace{-1mm}
\begin{equation}
\hat r^{(\mathrm s)} = \frac{1}{B}\sum_{b=1}^{B} R^{(\mathrm s)}_b,
\label{eq:emp_sensing_rate}
\end{equation}
with $R_b^{(\mathrm s)}$ evaluated for the corresponding clutter realization.

For the reference point, let $z_{\mathrm c}^{\star}$ and $z_{\mathrm s}^{\star}$ denote the communication and sensing reference values, and let $z_{\mathrm c}^{\min}$ and $z_{\mathrm s}^{\min}$ denote the lower normalization values that render two deficits dimensionless. The reference point is not required to be attainable. Both reference values are fixed hyperparameters and the values used are reported in Section~\ref{sec:results}. For a trade-off parameter $\mu\in[0,1]$, the normalized deficits are \looseness=-1
\begin{align}
d_{\mathrm c}=
\mu
\left[
\frac{z_{\mathrm c}^{\star}-\hat r^{(\mathrm c)}}
{z_{\mathrm c}^{\star}-z_{\mathrm c}^{\min}}
\right]_{+}, \,
d_{\mathrm s}=
(1-\mu)
\left[
\frac{z_{\mathrm s}^{\star}-\hat r^{(\mathrm s)}}
{z_{\mathrm s}^{\star}-z_{\mathrm s}^{\min}}
\right]_{+}.
\end{align}
A metric that already exceeds its reference thus contributes no deficit and the loss is governed by the remaining metric. The
GNN is trained to minimize the loss function
\begin{equation}
\mathcal L = \max\{d_{\mathrm c},d_{\mathrm s}\}. \label{eq:tcheby_loss}
\end{equation}
Finally, the GNN parameters are updated with the gradient $\nabla\mathcal L$, which is back-propagated through the network, as illustrated in Fig.~\ref{fig:method}.
%---------------------------------------------------------------------

%==============================================================================
\section{Simulation Results} \label{sec:results}
%==============================================================================
\begin{table}[t]
\centering
\caption{Main simulation and training parameters.}
\label{tab:sim_params}
\scriptsize
\setlength{\tabcolsep}{2pt}
\begin{tabular}{@{}llll@{}}
\toprule
\multicolumn{2}{@{}l}{\textbf{System parameters}} &
\multicolumn{2}{l}{\textbf{GNN and optimization parameters}} \\
\midrule
APs                & $M=3$                             & Layers            & $L=4$ \\
Antennas per ULA    & $N=4$                             & Hidden size       & $256$ \\
Users              & $K=12$                            & Mini-batch size   & $B=100$ \\
Sensing streams      & $S=|\mathcal{S}|=1$                         & Learning rate     & $2\times10^{-4}$ \\
Total Tx power     & $P_{\mathrm{tot}}=6$ & Training samples  & $20000$ \\
Per-AP power       & $P_m=2$               & Val.\ samples     & $2000$ \\            
Noise power        & $\sigma_k^2=0.1$      & Test samples      & $1000$ \\              
Sensing noise      & $\bar\varsigma^2=0.1$  & Trade-off weight  & $\mu\in\{0,0.25,\ldots,1\}$ \\           
Per-AP Tx SNR      & $13.0$ dB         & Comm.\ reference  & $z_{\mathrm c}^{\star}=0.65$  \\                 
Target gain        & $\zeta_m^2=1$     & Comm.\ lower      & $z_{\mathrm c}^{\min}=0$  \\ Nominal CSI error  & $\sigma_e=0.4$   & Sensing reference & $z_{\mathrm s}^{\star}=6.0$ \\
Nominal clutter    & $C=2$        & Sensing lower     & $z_{\mathrm s}^{\min}=0.9$ \\        
Rate quantile      & $\rho=0.05$        &Training clutter   & $C\in\{0,1,2,3,5\}$ \\
Samples/estimate   & $J=128$            &Training CSI range & $\sigma_e\sim\mathrm{Unif}[0.2,0.6]$ \\

\bottomrule
\end{tabular}
\end{table}
This section evaluates the proposed GNN framework under imperfect CSI and clutter-limited sensing. Unless stated otherwise, the nominal setup uses \(C=2\) clutter reflectors per AP and \(\sigma_e=0.4\), with the full parameter set listed in Table~\ref{tab:sim_params}. The robust communication rate denotes the empirical \(0.05\)-quantile of the minimum
user rate. All compared methods are evaluated on the same paired channel,
CSI error, and clutter realizations. The target and clutter angles are drawn independently and
uniformly per AP, $\phi_m,\phi_{m,c}\sim\mathrm{Unif}[0,\pi/2)$. Channels and clutter scenes are redrawn at
every training step, and no realization is seen twice. Each epoch therefore covers \(20000\) independent clutter scenes,
with the reflector count randomized per sample over \(C\in\{0,1,2,3,5\}\), while
the test set uses \(1000\) further independent scenes that are identical across
all compared methods. Training uses early stopping on a fixed validation set of \(2000\) samples.\looseness=-1

\begin{figure*}[t]
\centering
\begin{tikzpicture}
%================= (a) CSI sweep =================
\begin{axis}[
name=axa,
scale only axis, width=0.33\textwidth, height=0.115\textwidth,
xmin=0, xmax=0.4, xtick={0,0.1,0.2,0.3,0.4}, 
ymin=0, ymax=1.3, ytick={0,0.2,0.4,0.6,0.8,1.0,1.2},
grid=major, grid style={gray!24}, clip=true,
tick align=outside, xtick pos=bottom, ytick pos=left,
tick label style={font=\scriptsize}, label style={font=\footnotesize},
ylabel={\shortstack{Robust comm. rate\\(bit/s/Hz)}},
ylabel style={at={(axis description cs:-0.13,0.5)}, anchor=south},
xticklabels=\empty,
xtick style={draw=none},
]
% RZF
\addplot[color=blue!75!black, very thick, dashed, mark=square*, mark size=2.3pt, mark options={solid}] coordinates {(0,1.21498454) (0.1,0.64303988) (0.2,0.46467283) (0.3,0.36814243) (0.4,0.29726356)};
% Non-robust GNN, mu=0.75
\addplot[color=black, very thick, densely dotted, mark=o, mark size=2.3pt, mark options={solid, fill=white}]
coordinates {(0,0.52151513) (0.1,0.45449054) (0.2,0.39031577) (0.3,0.33113870) (0.4,0.28252393)};
\addplot[color=red!80!black, very thick, solid, mark=*, mark size=2.1pt]
coordinates {(0,0.50720519) (0.1,0.50994641) (0.2,0.49832106) (0.3,0.48910457) (0.4,0.44697523)};
\end{axis}

\begin{axis}[
name=axa2, at={(axa.south west)}, anchor=north west, yshift=-0.5cm,
scale only axis, width=0.33\textwidth, height=0.115\textwidth,
xmin=0, xmax=0.4, xtick={0,0.1,0.2,0.3,0.4},
ymin=0, ymax=6.4, ytick={0,2,4,6},
grid=major, grid style={gray!24}, clip=true,
tick align=outside, xtick pos=bottom, ytick pos=left,
tick label style={font=\scriptsize}, label style={font=\footnotesize},
xlabel={CSI-error variance $\sigma_e^2$},
ylabel={\shortstack{Mean sensing rate\\(bit/s/Hz)}},
ylabel style={at={(axis description cs:-0.13,0.5)}, anchor=south},
]
% Clutter-aware MVDR
\addplot[color=magenta!85!black, very thick, dashed, mark=triangle*, mark size=2.6pt, mark options={solid}]
coordinates {(0,4.19972897) (0.1,4.19362640) (0.2,4.19362640) (0.3,4.19362640) (0.4,4.19362640)};
% Non-robust GNN, mu=0.75
\addplot[color=black, very thick, densely dotted, mark=o, mark size=2.3pt, mark options={solid, fill=white}]
coordinates {(0,1.41064286) (0.1,1.40341878) (0.2,1.40313005) (0.3,1.40288651) (0.4,1.40266311)};
\addplot[color=red!80!black, very thick, solid, mark=*, mark size=2.1pt]
coordinates {(0,4.91932344) (0.1,4.45693731) (0.2,3.72496247) (0.3,2.41334772) (0.4,1.54076958)};
\end{axis}

%================= (b) clutter sweep =================
\begin{axis}[
    name=axb, at={(axa.north east)}, anchor=north west, xshift=1.7cm,
    scale only axis, width=0.33\textwidth, height=0.115\textwidth,
    xmin=0, xmax=5, xtick={0,1,2,3,4,5},
    ymin=0, ymax=0.75, ytick={0,0.2,0.4,0.6},
    grid=major, grid style={gray!24}, clip=true,
    tick align=outside, xtick pos=bottom, ytick pos=left,
    tick label style={font=\scriptsize}, label style={font=\footnotesize},
    xticklabels=\empty,
    xtick style={draw=none},
]
\addplot[color=blue!75!black, very thick, dashed, mark=square*, mark size=2.3pt, mark options={solid}] coordinates {(0,0.51888955) (1,0.51888955) (2,0.51888955) (3,0.51888955) (4,0.51888955) (5,0.51888955)};
\addplot[color=red!80!black, very thick, solid, mark=*, mark size=2.1pt] coordinates {(0,0.59068108) (1,0.51894534) (2,0.50548017) (3,0.50852412) (4,0.50600874) (5,0.52366)};
\addplot[
    color=black, very thick, densely dotted,
    mark=o, mark size=2.3pt,
    mark options={solid, fill=white}
] coordinates {
    (0,0.41290173) (1,0.41226962)
    (2,0.41093197) (3,0.41263506)
    (4,0.41244742) (5,0.41484475)
};
\end{axis}

\begin{axis}[
    name=axb2, at={(axb.south west)}, anchor=north west, yshift=-0.5cm,
    scale only axis, width=0.33\textwidth, height=0.115\textwidth,
    xmin=0, xmax=5, xtick={0,1,2,3,4,5},
    ymin=0, ymax=8.35, ytick={0,2,4,6,8},
    grid=major, grid style={gray!24}, clip=true,
    tick align=outside, xtick pos=bottom, ytick pos=left,
    tick label style={font=\scriptsize}, label style={font=\footnotesize},
    xlabel={Number of clutter reflectors $C$\vphantom{$\sigma_e$}},
]

\addplot[color=magenta!85!black, very thick, dashed, mark=triangle*, mark size=2.6pt, mark options={solid}] coordinates {(0,7.9128895) (1,5.857502) (2,4.1936264) (3,3.1101198) (4,2.4161386) (5,1.8906475)};
% Clean-trained GNN: sensing under increasing clutter
\addplot[ color=black, very thick, densely dotted, mark=o, mark size=2.3pt, mark options={solid, fill=white}] coordinates { (0,6.10715008) (1,2.35527134) (2,1.40323782) (3,1.00295770) (4,0.79343915) (5,0.64641470)};
\addplot[color=red!80!black, very thick, solid, mark=*, mark size=2.1pt] coordinates {(0,6.6954412) (1,5.4652839) (2,4.0561299) (3,2.8930261) (4,2.0793462) (5,1.458117)};
\end{axis}

%---- panel letters ----
\node[font=\footnotesize] at ([yshift=-1.15cm]axa2.south) {(a)};
\node[font=\footnotesize] at ([yshift=-1.15cm]axb2.south) {(b)};
\end{tikzpicture}

%---------------- shared legend ----------------
\begin{tikzpicture}
\begin{axis}[
    hide axis, xmin=0, xmax=1, ymin=0, ymax=1,
    scale only axis, width=1mm, height=1mm,
    legend columns=-1, legend cell align=left,
    legend style={draw=gray!45, font=\scriptsize,
                  at={(0.5,0.5)}, anchor=center,
                  /tikz/every even column/.append style={column sep=0.22cm}},
]
\addplot[color=red!80!black, very thick, solid, mark=*, mark size=2.1pt] coordinates {(0,0)};
\addlegendentry{Proposed GNN}
\addlegendimage{color=black, very thick, densely dotted,mark=o, mark size=2.3pt, mark options={solid, fill=white}}
\addlegendentry{Non-robust GNN}
\addplot[color=blue!75!black, very thick, dashed, mark=square*, mark size=2.3pt, mark options={solid}] coordinates {(0,0)};
\addlegendentry{RZF}
\addplot[color=magenta!85!black, very thick, dashed, mark=triangle*, mark size=2.6pt, mark options={solid}] coordinates {(0,0)};
\addlegendentry{MVDR}
\end{axis}
\end{tikzpicture}
\centering
\caption{Robust communication and sensing performance (a) under CSI uncertainty
and (b) versus the number of clutter reflectors.} \label{fig:resilience}
\end{figure*}
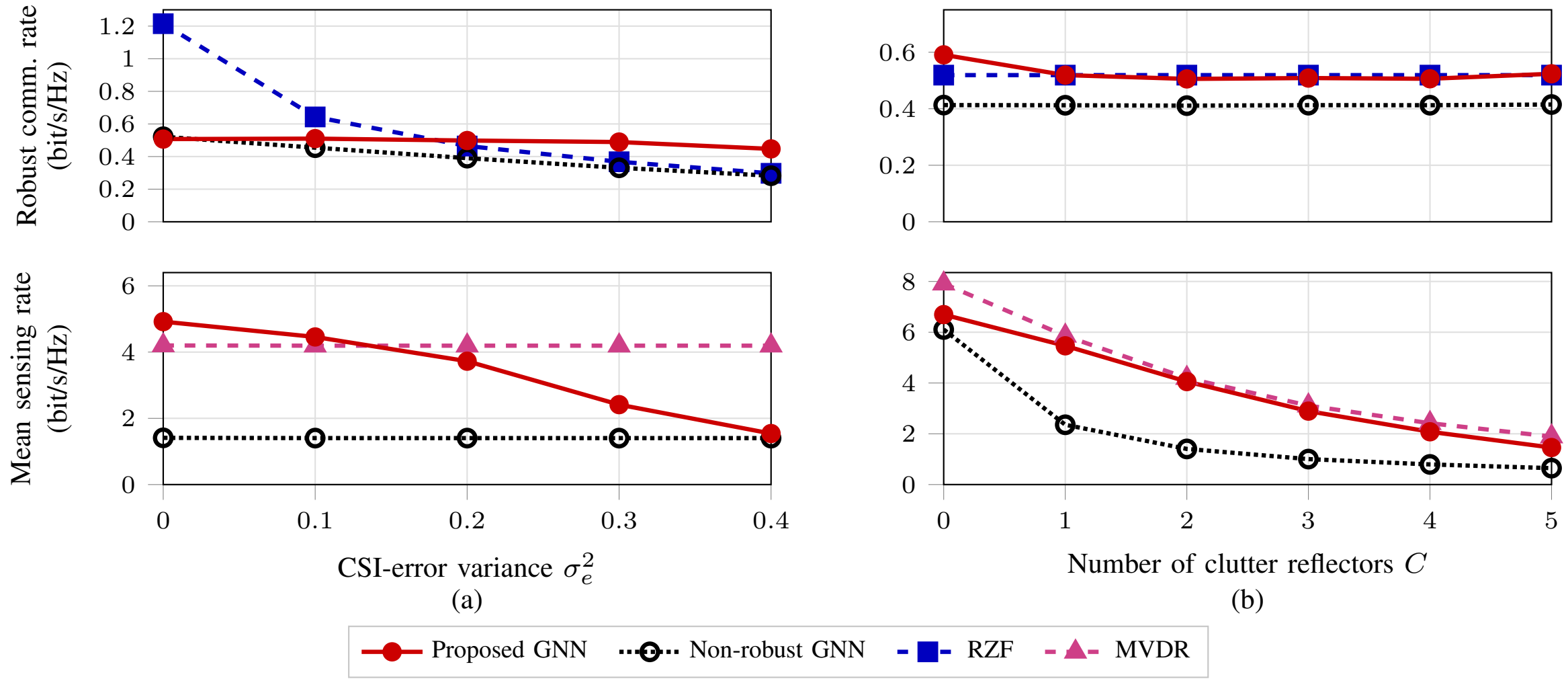

\textbf{Baselines.} Two analytical baselines are considered, each
dedicated to one functionality. The first baseline for communication utilizes a regularized zero-forcing (RZF) structure in~\cite{peel2005} computed from the stacked vector of channel estimates as 
\begin{equation}
\mathbf w_k^{(\mathrm{RZF})}
=
\left(
\min_{m\in\mathcal M}
\sqrt{\frac{P_m}{
\sum_{j\in\mathcal K}\|\bar{\mathbf w}_{mj}\|^2}}
\right)\bar{\mathbf w}_k,
\end{equation}
\begin{equation}
\bar{\mathbf w}_k
=
\left(
\sum_{j\in\mathcal K}
\tilde{\mathbf h}_j\tilde{\mathbf h}_j^H
+\lambda\mathbf I_{MN}
\right)^{-1}\tilde{\mathbf h}_k,
\end{equation}
with $\lambda=K\left(\sigma_k^2+\sigma_e^2P_{\mathrm{tot}}\right)$. The second baseline, dedicated to sensing, is a clutter-aware minimum variance distortionless response (MVDR) beamformer as follows~\cite{capon1969} 
\begin{equation}
\mathbf w_m^{(\mathrm{MVDR})}
=\sqrt{P_m}\,
\frac{(\mathbf Q_{\mathrm{cl},m}+\epsilon\mathbf I_N)^{-1}\mathbf a(\phi_m)}
{\bigl\|(\mathbf Q_{\mathrm{cl},m}+\epsilon\mathbf I_N)^{-1}\mathbf a(\phi_m)\bigr\|},
\end{equation}
where $\epsilon = 0.1$. For the resilience studies in Fig.~\ref{fig:resilience}, we additionally include a non-robust GNN trained with perfect CSI $\sigma_e=0$ and without clutter \(C=0\) utilizing the same architecture and a scalarization weight of $\mu=0.75$. This model is evaluated under the same test conditions as the proposed GNN.\looseness=-1

\renewcommand{\dbltopfraction}{0.9}
\renewcommand{\topfraction}{0.9}
\renewcommand{\textfraction}{0.07}
\renewcommand{\dblfloatpagefraction}{0.7}
\setlength{\textfloatsep}{8pt plus 2pt minus 2pt}
\setlength{\dbltextfloatsep}{8pt plus 2pt minus 2pt}
\begin{figure}[t]
    \centering
    \begin{tikzpicture}
\begin{axis}[
    width=1.02\columnwidth,
    height=0.58\columnwidth,
    xlabel={Robust comm. rate (bit/s/Hz)},
    ylabel={Mean sensing rate (bit/s/Hz)},
    xmin=0, xmax=0.72,
    xtick={0,0.1,0.2,0.3,0.4,0.5,0.6,0.7},
    ymin=0, ymax=6.5,
    ytick={0,1,2,3,4,5,6},
    grid=major,
    grid style={gray!25},
    tick align=outside,
    xtick pos=bottom,
    ytick pos=left,
    tick label style={font=\scriptsize},
    label style={font=\footnotesize},
    legend style={
        at={(0.0,0.0)},
        anchor=south west,
        legend columns=1,
        font=\tiny,
        cells={anchor=west},
        draw=gray!45,
        fill=white,
        inner sep=2pt,
        row sep=-1.5pt
    },
    clip=false
]
\addplot[only marks, color=red!80!black, mark=*, mark size=2.5pt]
    coordinates {(0,6.0360856) (0.38811663,5.6164918) (0.44943571,5.0075288) (0.50548017,4.0561299) (0.65924299,1.4139018)};
\addlegendentry{Proposed GNN}
\addplot[red!80!black, dashed, thick, no marks, forget plot]
    coordinates {(0,6.0360856) (0.38811663,5.6164918) (0.44943571,5.0075288) (0.50548017,4.0561299) (0.65924299,1.4139018)};
\addplot[only marks, color=blue!75!black, mark=square*, mark size=2.5pt] coordinates {(0.51888955,1.4200125)};
\addlegendentry{RZF}
\addplot[
    color=gray!75!black,
    dash dot,
    thick,
    no marks
]
coordinates {
    (0,4.1936264)          % MVDR sensing
    (0.51888955,1.4200125) % RZF
};
\addlegendentry{Time-sharing (MVDR--RZF)}
\addplot[only marks, color=magenta!85!black, mark=triangle*, mark size=3pt] coordinates {(0,4.1936264)};
\addlegendentry{MVDR}

\node[font=\tiny, anchor=west, xshift=4pt] at (axis cs:-0.015,5.48) {$\mu=0$};
\node[font=\tiny, anchor=south east, xshift=-3pt, yshift=3pt] at (axis cs:0.48811663,5.6164918) {$\mu=0.25$};
\node[font=\tiny, anchor=south west, xshift=3pt, yshift=3pt] at (axis cs:0.44943571,4.8775288) {$\mu=0.5$};
\node[font=\tiny, anchor=south west, xshift=3pt, yshift=3pt] at (axis cs:0.50548017,3.7561299) {$\mu=0.75$};
\node[font=\tiny, anchor=south east, xshift=-3pt, yshift=3pt] at (axis cs:0.70924299,0.3039018) {$\mu=1$};
\end{axis}
\end{tikzpicture}
    \vspace{-3mm}
    \caption{Communication--sensing operating plane.}
    \label{fig:alpha_tradeoff}
\end{figure}
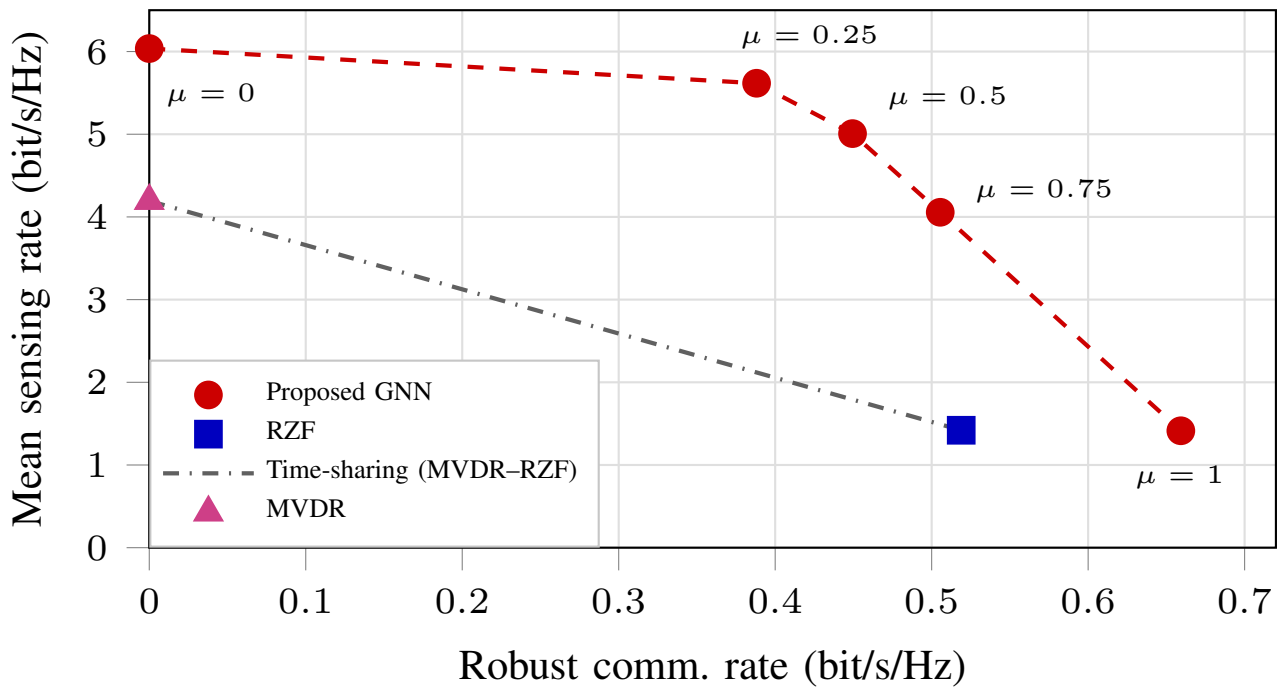

\subsection{Communication-Sensing Trade-off} \label{sec:tradeoff}
The benefit of jointly optimizing robust communication and sensing performance is illustrated in Fig.~\ref{fig:alpha_tradeoff}. When only optimizing for sensing by setting \(\mu=0\), a mean sensing rate of \(6.036\) bit/s/Hz is reached, exceeding the clutter-aware MVDR baseline by \(43.9\%\). The MVDR beamformer maximizes the ratio between target gain and clutter leakage at each AP separately, whereas the joint SCNR in~\eqref{eq:scnr} combines the contributions of all APs in a single ratio. The GNN optimizes this joint ratio directly and therefore attains higher sensing rates. When only optimizing communication by setting \(\mu=1\), a robust communication rate of \(0.659\)~bit/s/Hz is reached, which exceeds RZF by \(27.0\%\) at a comparable sensing rate of \(1.414\)~bit/s/Hz. RZF is computed from the channel estimates without accounting for the CSI error, whereas the GNN is trained on the outage quantile. In both cases, the remaining sensing rate stems from the incidental target illumination of the communication beams. The intermediate models reveal the trade-off between communication reliability and sensing performance.  At \(\mu=0.75\), the proposed method retains \(97.4\%\) of the RZF robust communication rate while achieving \(96.7\%\) of the MVDR sensing rate. This point is therefore selected for the resilience studies below. The dash-dotted line corresponds to time sharing between RZF and MVDR, where each baseline is active for a fraction of the time. All learned operating points lie above this line, which shows that the GNN reaches trade-offs that switching between the two baselines cannot achieve. \looseness=-1

\subsection{Resilience to CSI Uncertainty and Clutter}
This subsection examines whether the operating point selected in Section~\ref{sec:tradeoff} remains reliable when the impairments deviate from the nominal setup, including CSI-error levels and clutter counts not seen during training. The proposed GNN with $\mu=0.75$ is evaluated at \(C=2\) over different CSI-error levels in Fig.~\ref{fig:resilience}(a). At perfect CSI, RZF achieves the highest robust communication rate, followed by the non-robust GNN. However, the proposed GNN outperforms the non-robust GNN in both communication and sensing at every tested positive variance between $0.1$ and $0.4$. At $\sigma_e^2=0.2$, the proposed GNN achieves approximately $28\%$ higher robust rate than the non-robust model, while providing $3.725$ versus $1.403$ bit/s/Hz sensing rate. The robust model also surpasses RZF at all tested variances from $0.2$ onward. For sensing, the proposed GNN exceeds the clutter-aware MVDR baseline at $\sigma_e^2=0$ and $0.1$, and retains approximately $88.8\%$ of the MVDR sensing rate at $\sigma_e^2=0.2$, while simultaneously serving communication users. Training covers standard deviations $\sigma_e\in[0.2,0.6]$, corresponding to variances $\sigma_e^2\in[0.04,0.36]$. Thus, the evaluation includes error levels outside the training range. At the unseen variance $\sigma_e^2=0.4$, which is $2.5$ times the nominal variance, the proposed GNN achieves $0.447$ bit/s/Hz robust communication rate, exceeding RZF and the non-robust GNN by $50.4\%$ and $58.2\%$, respectively. Since the impairment levels are inputs to the GNN, it adapts its beamformers to this unforeseen condition, which corresponds to the adaptation capability of resilience. Nevertheless, its sensing rate decreases as the CSI error grows, because the communication deficit increasingly dominates the Tchebycheff loss and shifts resources toward communication.\looseness=-1

Fig.~\ref{fig:resilience}(b) studies the effect of the number of clutter reflectors at \(\sigma_e=0.4\). For $C=1,\ldots,5$, the proposed GNN provides approximately $23\%$ to $26\%$ higher robust communication rates than the non-robust GNN. Additionally, the proposed model outperforms the non-robust one in sensing throughout the sweep. At \(C=4\), a clutter count never presented during training, the proposed model achieves \(2.079\) bit/s/Hz sensing rate, approximately $86.1\%$ of the MVDR rate, while still providing \(0.506\) bit/s/Hz robust communication rate. The clutter-covariance input therefore allows the GNN to adapt its beamformers to clutter densities of varying size, including an unseen one, while the Tchebycheff objective prevents the solution from collapsing into a pure sensing design. This result demonstrates resilience to an unseen intermediate clutter count, while maintaining both communication and sensing functionality.\looseness=-1
%==============================================================================
\section{Conclusion}
%==============================================================================
This work proposes an impairment-aware heterogeneous GNN for cell-free ISAC beamforming
under imperfect CSI and sensing clutter. A single jointly learned operating point approaches the respective rates of the two baselines, whereas training under ideal conditions provides weaker performance under the tested impairments. The evaluation at an unseen clutter count and unseen CSI error variances provides evidence of resilience. However, the sensing rate of the proposed GNN decreases as the CSI uncertainty increases, although its robust communication rate remains above that of RZF and the non-robust GNN at large CSI error levels. This motivates future designs with an explicit sensing rate constraint alongside the communication reliability objective.

%==============================================================================

\bibliographystyle{IEEEtran}
\bibliography{main}

\end{document}